\documentclass[nofootinbib,preprint]{revtex4-1}
\usepackage[T1]{fontenc}
\usepackage{graphicx}
\usepackage{rotating}
\usepackage{bm}        
\usepackage{amssymb}   
\usepackage{bm}
\usepackage{float}
\usepackage{tikz}
\usepackage{diagbox}
\usepackage{braket}
\usepackage{color} 
\usepackage{colortbl}
\usepackage{amsmath}

\def\etal{{\it et al.}}

\def\jpca#1#2#3{{\it J.~Phys.~Chem.~{\rm A}}~{\bf #1},\ {#2}\ (#3)}
\def\jpcc#1#2#3{{\it J.~Phys.~Chem.~{\rm C}}~{\bf #1},\ {#2}\ (#3)}
\def\jcp#1#2#3{{\it J.~Chem.~Phys.}~{\bf #1},\ {#2}\  (#3)}

\def\prl#1#2#3{{\it Phys.~Rev.~Lett.}~{\bf #1},\ {#2}\ (#3)}
\def\sci#1#2#3{{\it Science}~{\bf #1},\ {#2}\ (#3)}
\def\nat#1#2#3{{\it Nature}~{\bf #1},\ {#2}\ (#3)}
\def\natc#1#2#3{{\it Nat. Commun.}~{\bf #1},\ {#2}\ (#3)}
\def\mp#1#2#3{{\it Mol. Phys.}~{\bf #1},\ {#2}\ (#3)}
\def\jpb#1#2#3{{\it J. Phys. B: At. Mol. Opt. Phys.} {\bf #1},\ {#2}\ (#3)}

\def\pccp#1#2#3{{\it Phys. Chem. Chem. Phys.}~{\bf #1},\ { #2}\ (#3)}

\def\jctc#1#2#3{{\it J. Chem. Theor. Comp.}~{\bf #1},\ {#2}\ (#3)}
\def\ijqc#1#2#3{{\it Int. J. Quant. Chem.}~{\bf #1},\ {#2}\ (#3)}

\newcommand{\req}[1]{\begin{align}#1\end{align}}

\makeatletter

    \def\CT@@do@color{%
      \global\let\CT@do@color\relax
            \@tempdima\wd\z@
            \advance\@tempdima\@tempdimb
            \advance\@tempdima\@tempdimc
    \advance\@tempdimb\tabcolsep
    \advance\@tempdimc\tabcolsep
    \advance\@tempdima2\tabcolsep
            \kern-\@tempdimb
            \leaders\vrule
                    \hskip\@tempdima\@plus  1fill
            \kern-\@tempdimc
            \hskip-\wd\z@ \@plus -1fill }
    \makeatother

\def\k1{k_1}
\def\k2{k_2}
\def\q1{q_1}
\def\q2{q_2}

\def\({\left (}
\def\){\right )}
\def\[{\left [}
\def\]{\right ]}

\newcommand{\beq}{\begin{equation}}
\newcommand{\eeq}{\end{equation}}

\DeclareMathAlphabet\mathbfcal{OMS}{cmsy}{b}{n}

\begin{document}
\date{\today}
\flushbottom \draft
\title{Quantum Gaussian process model of potential energy surface for a polyatomic molecule
}
\author{J. Dai and R. V. Krems}
\affiliation{
Department of Chemistry, University of British Columbia, Vancouver, B.C. V6T 1Z1, Canada \\
Stewart Blusson Quantum Matter Institute, Vancouver, B.C. V6T 1Z4, Canada}

\begin{abstract}
With gates of a quantum computer designed to encode multi-dimensional vectors, 
projections of quantum computer states onto specific qubit states can produce kernels of reproducing kernel Hilbert spaces. 
We show that quantum kernels obtained with a fixed ansatz implementable on current quantum computers can be used for accurate regression models of global potential energy surfaces (PES) for polyatomic molecules. 
To obtain accurate regression models, we apply Bayesian optimization to maximize marginal likelihood by varying the parameters of the quantum gates. 
This yields Gaussian process models with quantum kernels. We illustrate the effect of qubit entanglement in the quantum kernels and explore the generalization performance of 
quantum Gaussian processes by extrapolating global six-dimensional PES in the energy domain. 

\end{abstract}

\maketitle

\section{Introduction}

Predicting properties of complex molecules from first principles is considered to be one of the most promising applications of quantum computing. 
A computation of molecular properties within the Born-Oppenheimer approximation requires solving the electronic structure problem, fitting the results of potential energy calculations to produce global PES and solving the nuclear dynamics problem with the PES thus obtained. 
Several algorithms have been recently developed for solving electronic structure \cite{electronic-structure,electronic-structure-1,electronic-structure-2,electronic-structure-3,electronic-structure-4,electronic-structure-5,electronic-structure-6,electronic-structure-7,electronic-structure-8,electronic-structure-9,electronic-structure-10,electronic-structure-11,electronic-structure-12,electronic-structure-13,electronic-structure-14,electronic-structure-15} and nuclear dynamics \cite{nuclear-dynamics,nuclear-dynamics-1,nuclear-dynamics-2,nuclear-dynamics-3,nuclear-dynamics-4,nuclear-dynamics-5} problems on noisy intermediate-scale quantum (NISQ) computers. 
However, quantum algorithms for producing global PES of polyatomic molecules have not yet been demonstrated. 
The present work builds a quantum regression model of a six-dimensional PES for the molecular ion H$_3$O$^+$.  
Our results demonstrate a comparison with the corresponding classical models and illustrate the role of entanglement of the qubits used for the quantum algorithm of constructing PES.

Recent work has demonstrated that PES of polyatomic molecules can be accurately represented by machine learning (ML) regression models, based on neural networks  \cite{ML-for-PES2, ML-for-PES3, NNs-for-PES, NNs-for-PESa, NNs-for-PES-1a, NNs-for-PES-1b, NNs-for-PES-1c, NNs-for-PES-2, NNs-for-PES-3, NNs-for-PES-4, NNs-for-PES-5, NNs-for-PES-6, carrington, meuwly} 
or kernel methods \cite{meuwly, gp-1, gp-2, gp-3, jie-jpb, gp-for-PES-2, gp-for-PES-3, gp-for-PES-4, gp-for-PES-5, gp-for-PES-6, gp-for-PES-7, gp-for-PES-8, gp-for-PES-9, gp-for-PES-10, gp-for-PES-11, gp-for-PES-12, kernel-for-PES, rabitz-1, rabitz-2, rabitz-3, unke}. 
Quantum computers have opened the possibility to research the quantum analogues of ML algorithms  \cite{qml,qml1,vcc,vcc-1,vcc-2,qK-SVM,qK-SVM-1,qK-SVM-2,qK-SVM-3,qK-SVM-4,qK-SVM-5, qK-SVM-6, quantum-regression,quantum-regression-1,quantum-regression-2,quantum-regression-3,qgp}. 
It has been shown that gate-based quantum devices can be used to build quantum kernels for kernel ML models \cite{qml,qml1, qK-SVM,qK-SVM-1,qK-SVM-2,qK-SVM-3,qK-SVM-4,qK-SVM-5, qK-SVM-6, qgp}. 
While most applications of quantum kernels have been for support vector classification of low-dimensional data \cite{qml1,qK-SVM,qK-SVM-1,qK-SVM-2,qK-SVM-3,qK-SVM-4,qK-SVM-5, qK-SVM-6}, 
several studies have considered quantum algorithms for regression \cite{quantum-regression, quantum-regression-1, quantum-regression-2, quantum-regression-2, quantum-regression-3, qgp}.  
Particularly relevant for the present work is Ref. \cite{qgp} that applied Gaussian process regression to several model applications, such as regression of the one-dimensional function $x \sin x$. 
 The goal of Ref. \cite{qgp} was to simulate classical kernels using coherent states, or truncations of coherent states. 
In order to extend this work to regression problems for fitting PES, it is necessary to overcome several challenges. 
First, there is no general quantum circuit ansatz for building performant quantum kernels for PES interpolation. 
It is not known how to build a sequence of quantum gates in order to build the best quantum kernel for accurate models of PES. 
Second, accurate kernel regression models for complex problems with sparse data require optimization of kernel parameters. 
However, quantum kernel estimation is expensive, requiring many quantum measurements for each pair of training points. 
In addition, as will be illustrated in this work, the cost function used to train regression models
with quantum kernels can be very sensitive to quantum circuit parameters. 
This makes kernel parameter optimization difficult. 
Third, the number of quantum kernel parameters grows quickly with the number of qubits and gates in the corresponding quantum circuit. 
This precludes grid search of optimal kernel parameters, often used for building classical kernel ridge regression models. This also makes search of optimal quantum circuit ansatz difficult. 

Here, we demonstrate that quantum kernel regression models with a fixed quantum circuit ansatz readily deployable on current gate-based quantum computers can yield comparable accuracy with classical ML models.  Our focus is on building accurate models with a small number of training points, aiming to produce global PES with a small number of {\it ab initio} potential energy calculations.  
To achieve this, we employ Bayesian optimization for tuning the parameters of quantum gates and optimize kernels by maximizing a modified version of  log marginal likelihood. 
We consider two problems: interpolation of PES in a six-dimensional (6D) configuration space and extrapolation of PES in the energy domain. We show that the quantum models may exhibit better extrapolation accuracy than classical models 
with radial basis function kernels \cite{universality-of-RBF-kernels}, when trained by the same number and distribution of potential energy points. We also show that the accuracy of quantum models is significantly enhanced by two-qubit gates, which illustrates a  critical role
of qubit entanglement in quantum kernels for regression problems. 
By demonstrating Bayesian regression models with quantum kernels, our work complements Ref. \cite{qgp} to set the stage for the quantum analogue of Bayesian optimization on quantum computing devices.

\section{Classical vs quantum models}

We use Gaussian process (GP) models to represent PES of the molecule H$_3$O$^+$. The molecular geometry is described by the six-dimensional (6D) vector $\bm x$, as in our previous work \cite{jun-paper}, where we built classical GP models of PES for H$_3$O$^+$.
A GP model is trained by $n$ input - output pairs, with inputs represented by $n$ molecular geometries $\bm x_i$ and outputs by $n$ corresponding values of the potential energy, collected into a column vector $\bm y$.
The prediction of potential energy at an arbitrary point $\bm x^\ast$ in the 6D input space is given by \cite{gp-book}: 

\begin{eqnarray}
\hat{f}(\bm x^\ast) 
= {\bm k}^\top(\bm x^\ast)   \left [ {\bf K} + \sigma^2 {\bf I} \right ]^{-1} \bm y
\label{gp-prediction}
\end{eqnarray}
where $\sigma^2$ is a hyperparameter representing variance of data noise, $\bf I$ is the identity matrix, $\bf K$ is an $n \times n$ kernel matrix with entries $k(\bm x_i, \bm x_j)$, ${\bm k}^\top(\bm x^\ast)$ is the transpose of a column vector with $n$ entries
$k(\bm x^\ast, \bm x_i)$, and $\bm x_i$ and $\bm x_j$ represent the molecular geometries for the training points $i$ and $j$.  Because PESs are noiseless, we set $\sigma^2$ to zero. 

The function $k(\bm x, \bm x ')$ yielding the elements of the kernel matrix is the covariance function of the GP \cite{gp-book}. 
It must satisfy the properties of a kernel function of a reproducing kernel Hilbert space (RKHS). Specifically, $k(\bm x, \bm x ')$ must be positive-definite and symmetric to interchange of $\bm x$ and $\bm x'$. 
In the present work, we build GP models with classical and quantum kernels. In both cases, the prediction of the model is given  by Eq. (\ref{gp-prediction}). The difference is in the kernel matrix $\bf K$. 
For classical models, we use radial basis functions (RBF) as the kernel function,
\begin{eqnarray}
k(\bm x, \bm x') = \exp\left (  - \theta || \bm x - \bm x'||^2 \right ).
\end{eqnarray}
RBF kernels are known to be universal \cite{universality-of-RBF-kernels} and provide benchmark results for quantum models developed in this work. 

For quantum kernels, we consider a quantum computer with $m$ qubits,  initially in state $|0^m\rangle$. 
A sequence of gates operating on these qubits produces a quantum state ${\cal U}(\bm x)| 0^m \rangle$. 
The  measurable square of the inner product
\begin{eqnarray}
k(\bm x, \bm x') =  |\langle 0^m | {\cal U}^\dagger(\bm x') {\cal U}(\bm x) | 0^m \rangle|^2
\label{quantum-kernels}
\end{eqnarray}
satisfies all the properties of a kernel of an RKHS.  In order to build such quantum kernels, one must encode information about input vectors into parameters of the quantum gates of a quantum computer.

In the present work, we use the quantum circuit depicted in Figure 1 to build quantum kernels. 
This quantum circuit was introduced in Ref. \cite{qK-SVM-2} for classification problems. 
We use one qubit to represent one dimension of the input space, resulting in a 6-qubit quantum circuit for the present problem. Each qubit is initialized in state $|0 \rangle$. 
Following the initialization, quantum states are created by a sequence of gate operations ${\cal U}^\dagger(\bm x') {\cal U}(\bm x)$, as depicted in the upper panel of Figure 1. The values of the kernels are obtained by projecting the resulting quantum states onto the state 
$|0^{m} \rangle$. 

As shown in Figure 1, the unitary transformation $\cal U$ includes a sequence of three types of quantum gates: the Hadamard gates ($H$), 
\req{
H
=
\frac{1}{\sqrt{2}}
  \begin{pmatrix}
 1&
   1 \\
   1 &
   -1 
   \end{pmatrix}
 }
which put the individual qubits into coherent superposition states, the single-qubit rotation gates $R_Z$,
\req{
R_Z(\phi_i)
=
  \begin{pmatrix}
 e^{-i \phi_i} &
   0 \\
   0 &
   e^{i \phi_i} 
   \end{pmatrix}
 }
\noindent
and the two-qubit rotation gates $R_{ZZ}$,
  \req{
R_{ZZ}(\phi_{ij}) =
  \begin{pmatrix}
 e^{-i \phi_{ij}}~& 0& 0& 0 \\
 0 & e^{i \phi_{ij}}~  & 0  & 0 \\
 0 & 0  & e^{i \phi_{ij}}  & 0 \\
 0 & 0  & 0 & e^{-i \phi_{ij}} \\
\end{pmatrix}.
 }
The two-qubit gates introduce entanglement. 

The input vectors $\bm x$ are encoded into the quantum gates as follows: 
\req{
\phi_{i} = \bm x^i/ \theta_i
\\
\phi_{ij} = \exp (-{(\bm x^i-\bm x^j)}/{\theta_{ij}}),
}
where the superscripts in $\bm x^i$ and $\bm x^j$ denote the $i$-th and  $j$-th components of the 6D vector $\bm x^\top = \left [ \bm x^1, \dots, \bm x^6 \right ]$, and 
$\theta_i$ and $\theta_{ij}$ are parameters of the quantum circuit to be optimized. As shown in Figure 1, the unitary transformation $\cal U$ is built as 
\req{
\mathcal{U}= {\bf U} {H}^{\otimes n} {\bf U} {H}^{\otimes n},
}
with 
\begin{equation}
U= \exp \left [-i \left (\sum_i^m\phi_{i}(\bm x,\theta_i) {\sigma}_{Z,i} +\sum^m_{i,j > i} \phi_{ij}(\bm x,\theta_{ij}) {\sigma}_{Z,i} {\sigma}_{Z,j}  \right)\right ]
\label{circuit}
\end{equation}
where $ \sigma_{Z,i}$ is the Pauli $Z$-gate acting on qubit $i$, and the second term in the exponent correspond to the ${R}_{ZZ}$ gate.
This ansatz includes a sequence of two-qubit rotations, entangling each pair of the qubits in the circuit.
The order of the individual $R_{ZZ}$ gates in Eq. (\ref{circuit}) is arbitrary, because the $ \sigma_{Z,i}$ operators commute. 
The parameters $\theta_i$ are independent for each one-qubit rotation gate in $\bf U$. In order to simplify the optimization of the kernel parameters, we require that the parameters of all two-qubit gates $\theta_{ij}$ be the same and set them equal to a single variable parameter $\theta_{12}$.
 The number of free parameters in the quantum kernel is thus equal to the number of $R_Z$ gates plus one, for a total of 7 parameters in $\bf U$.

The covariance functions of the GP models are thus parametrized by $\theta$ in the classical models and $\theta_{i=1,\dots,6}$ and $\theta_{12}$, hereafter represented collectively by $\bm \theta$, in the quantum models. 
GP models are trained by maximizing the logarithm of marginal likelihood (LML), which yields optimal parameters of the kernels \cite{gp-book}. For GPs, LML can be written in closed form in terms of the kernel matrix $\bf K$ and its determinant as follows \cite{gp-book}: 
\req{
\log {\cal L}(\bm \theta)  = -\frac{1}{2} \bm y^\top \left ( {\bf K} + \sigma^2 {\bf I} \right )^{-1} \bm y - \frac{1}{2} \log | {\bf K} + \sigma^2 {\bf I }| - \frac{n}{2} \log{2 \pi},
\label{LML}
} 
where the dependence on $\bm \theta$ is through the elements of the kernel matrix. While it is straightforward to train classical GP models with the RBF kernel by maximizing LML, it will be illustrated in the next section that LML for quantum models is extremely sensitive to $\bm \theta$ in some parts of the parameter space, leading to rapid variation of LML and lack of convergence of LML optimization. In order to overcome this problem, we show that quantum models can be trained by optimizing the following objective function instead of the LML: 
\req{
{\cal O}(\bm \theta) = \log [{\cal L}(\bm \theta) + a]
\label{o-function}
}
where $a$ is a hyperparameter, set to 1 in the present work. It will be shown that the constant $a$ stabilizes optimization of LML and improves convergence.

To build quantum kernels, we use simulated qubits as implemented in the IBM qiskit package, using Statevector \cite{qiskit}. 
Quantum states are generated by the operation of gate sequences on qubits initially all in state $| 0 \rangle$. The gate operations are noiseless. 
The kernels as defined in Eq. (\ref{quantum-kernels}) are computed from the corresponding probability amplitudes in the quantum states of $m$ qubits after the sequence of gate operations. 
In order to examine the role of qubit entanglement, we consider two types of kernels for the quantum models: (a) kernels constructed as described above with 7 parameters $\theta_{i}$ and $\theta_{12}$; (b) kernels constructed as described above, but with all two-qubit gates $R_{ZZ}$ replaced with identity matrices, yielding quantum circuits with 6 free parameters and no entanglement between qubits. We will refer to these kernels as entangled and unentangled kernels, respectively.

\section{Results}

Although Ref. \cite{qK-SVM-2} illustrated that the quantum circuit ansatz described in the previous section can be used to build kernels for classification problems, this ansatz has not been used for regression models. 
Therefore, out first goal is to explore the possibility of using the quantum circuit depicted in Figure 1 for regression problems. Specifically, we aim to build accurate interpolation and extrapolation models with a limited number of training points (200 to 1500 for a 6D problem). 
In this limit, and especially for extrapolation problems, kernel regression models must be sensitive to kernels. We use a comparison with the models based on optimized RBF kernels to benchmark the performance of the quantum kernels. 
It should be noted that RBF kernels do not always represent the best classical kernels for kernel models of PES. As we demonstrated previously, classical GP models of PES can be improved by increasing the kernel complexity by combining different simple mathematical forms of kernels into composite kernels \cite{jun-paper, kasra}. However, RBF kernels are proven to be universal \cite{universality-of-RBF-kernels} and represent one of the most frequently used type of kernels. 
Our goal is not to illustrate that quantum kernels can outperform classical kernels for small data regression problems. 
Rather, we aim to show that quantum kernels can produce regression models of similar accuracy as classical kernels.

Specifically, the present section illustrates: 

\begin{itemize}

\item how to optimize quantum circuits to build accurate quantum GP models; 

\item the feasibility of building accurate GP regression models with quantum kernels using a fixed quantum circuit ansatz depicted in Figure 1; 

\item comparison of quantum GP models for interpolation and extrapolation (in the energy domain) of PES with the classical models with optimized RBF kernels;

\item comparison of quantum GP models of PES with and without entanglement between qubits.

\end{itemize}

The {\it ab initio} results for the PES of H$_3$O$^+$ are taken from Ref. \cite{h3o+}. 
There are a total of 31124 potential energy points, spanning the energy range $[0, 21000]$ cm$^{-1}$. We construct global 6D PES by training GP models using $n$ {\it ab initio} points in a specific energy interval. 
The value of $n$ and the energy range for the training points is specified in the caption for each figure. The accuracy of the resulting models is quantified by computing the root mean squared error (RMSE) 
\begin{eqnarray}
\textrm{RMSE} = \sqrt{\frac{1}{N} \sum_{i=1}^{N} \left( y_i - \hat{f}(\bm x_i) \right)^2},
\label{RMSE}
\end{eqnarray}
where $\hat{f}$ are the GP model predictions given by Eq. (\ref{gp-prediction}), $y_i$ represent the {\it ab initio} potential energy points from Ref \cite{h3o+}, and the sum extends 
over 
all {\it ab initio} points that are not used for training the models. For models trained by potential energy points from a limited range of energies (e.g. at energies $\leq 10,000$ cm$^{-1}$), these RMSEs covering the entire energy range up to 21,000 cm$^{-1}$ quantify the ability of GP models to extrapolate in the energy domain.

\subsection{Quantum kernel optimization}

For classical GP models with simple analytical kernel functions, LML optimization is usually performed with a gradient-based optimization method, quickly converging to desired estimates of kernel parameters. 
As follows from the above description of quantum kernels, LML maximization for quantum models requires optimization of a large number of parameters $\bm \theta$, with kernels given by the probability amplitudes in a quantum state. 
When implemented on a quantum computer, the present algorithm will yield kernels as quantum measurement outcomes, instead of analytical functions. 
This makes optimization of LML, or equivalently ${\cal O}(\bm \theta)$, much more challenging.  
In this section, we illustrate that accurate quantum GP models can be obtained by optimizing ${\cal O}(\bm \theta)$ in Eq. (\ref{o-function}) with Bayesian optimization (BO).

BO is a gradient-free optimization method that uses a balance between the prediction of a GP and the Bayesian uncertainty of the prediction to determine how to sample the function under optimization \cite{rodrigo-bo}. Here, we apply BO to find the parameters of the quantum circuit $\bm \theta$ that maximize ${\cal O}(\bm \theta)$. BO begins with the evaluation of ${\cal O}(\bm \theta)$ at a small number of randomly selected values of $\bm \theta$. The results of these evaluations are used to train a (classical) GP model ${\cal F}(\bm \theta)$ characterized by the mean of the GP $\cal F$ denoted as $\mu(\bm \theta)$ and by the uncertainty of the GP $\cal F$ denoted as $\sigma(\bm \theta)$. The subsequent evaluation of ${\cal O}(\bm \theta)$ is performed at the maximum of the acquisition function $\alpha(\bm \theta)$ defined  as 
\req{
\alpha(\bm \theta) = \mu(\bm \theta) + \kappa \sigma(\bm \theta),
\label{aw}
}
 where $\kappa$ is a hyperparameter that determines the balance between exploration and exploitation.  
The result of the new evaluation of ${\cal O}(\bm \theta)$ is added to the set of the previous evaluations and the new set of $\cal O$ values is used to train a new GP model $\cal F$. 
The procedure is iterated until convergence is reached. 

We use RBF kernels for the GP models $\cal F$, initialize BO with 20 randomly chosen points and typically reach optimal results with $\sim 30 - 100$ iterations sampling 6 or 7 dimensions of the $\bm \theta$ parameter space. We use the value of 
$\kappa$ in Eq. (\ref{aw}) set to $1$. We have repeated calculations with multiple values of $\kappa$ and found that this choice of $\kappa$ leads to optimal convergence of BO for the present problems.

Figure 2 illustrates the results of optimization of LML using the objective functions defined in Eqs. (\ref{LML}) and (\ref{o-function}) for unentangled (left panel) and entangled (right panel) kernels. 
These optimization problems vary 6 and 7 parameters, respectively. 
LML exhibits sharp variation with $\bm \theta$, with characteristic drops (c.f., right panel of Figure 2), suggesting 
 the presence of singularities for some values of the quantum circuit parameters. 
Qubit entanglement makes the optimization of LML more challenging.
However, introducing a constant under the logarithm of the objective function as in Eq. (\ref{o-function}) stabilizes optimization and improves convergence 
for GP models with both unentangled and entangled kernels. We have repeated optimization with several different values of $a \in [0.1, 10]$  in Eq. (\ref{o-function})  and found that the results are not sensitive to the value of $a$. All of the calculations reported in this work use the value $a = 1$.

Figure 3 illustrates the effect of qubit entanglement on the results of LML optimization and the accuracy of the corresponding quantum GP models quantified by the RMSE over the entire data set. 
The results illustrate that including qubit entanglement enhances the accuracy of the quantum models. The right panel of Figure 3 shows that accurate models of 6D PES based on entangled kernels 
can be obtained with as few as 20 iterations of kernel optimization. This illustrates both the feasibility of obtaining accurate regression models with the fixed ansatz in Figure 1, and the efficiency of BO for optimizing quantum circuits for quantum regression problems.

Figure 4 illustrates convergence of BO of LML for quantum models with entangled kernels based on different numbers of training points. As expected, LML increases and RMSE decreases with the number of training points. 
The optimization of LML converges with less than 30 iterations of BO for all three models.  Figure 4 illustrates that quantum GPs produce reasonable models of 6D PES, when trained with as few as 200 potential energy points randomly sampled from the 6D configuration space.  
It can be observed that the optimization of the quantum circuit parameters reduces the RMSE for models with 1000 potential energy points by a factor of 3. 
These results illustrate that the quantum circuit ansatz introduced in Ref. \cite{qK-SVM-2} for classification problems 
is also effective for regression problems and that it is flexible enough to allow learning of complex functions by optimization of quantum gate parameters.

\subsection{Quantum vs classical GP models of PES -- interpolation}

Figure 5 illustrates the interpolation performance of the optimized quantum regression model of the 6D PES of H$_3$O$^+$ built with 1000 potential energy points. 
The line represents the quantum model predictions and the symbols --   the potential energy points randomly sampled as functions of the separation $R$ between the centers of mass of H$_2^+$ and OH fragments.
At each value of $R$, we locate the energy point in the original set of {\it ab initio} points by varying the angles and/or the interatomic distances within the fragments. This energy point is then compared with the GP predictions.
The training data for this model are sampled from the entire energy range of the PES. The quantum model is based on the entangled kernel and is obtained with 72 iterations of BO. The RMSE of the model is 82.30 cm$^{-1}$. 
While this is a remarkable performance of the quantum kernel, we note that the accuracy of the model can be further increased by increasing the number of training points (c.f., Figure 4).

It is instructive to compare the performance of this quantum model with that of the quantum model based on unentangled qubits and of the classical GP model. Figure 6 shows that the quantum model with entangled qubits 
is significantly more accurate than the quantum model with unentangled qubits. This illustrates the importance of two-qubit gates in the quantum circuit ansatz. 
Figure 6 also illustrates that the accuracy of the GP model with the optimized RBF kernel is very close to the accuracy of the model with the entangled kernel, except for $n=100$. Both models approach the RMSE of about $37$ cm$^{-1}$ as the number of training points increases.

\subsection{Extrapolation in the energy domain}

Several recent studies have explored the application of GP models for extrapolation problems. It was shown that the generalization accuracy of GP models increases if the complexity of GP kernels is increased by combining different simple kernels into composite kernels through an algorithm using Bayesian Information Criterion as the model selection metric \cite{bic,extrapolation-1,extrapolation-2}. It was shown that GP models thus constructed 
can extrapolate the properties of complex quantum systems across quantum phase transition lines \cite{extrapolation-3}. 
The same approach was used to enhance the accuracy of GP models of PES for polyatomic molecules \cite{jun-paper,hiroki,kasra}.
Since quantum circuits offer a conceptually different approach to building kernels for GP models, it is instructive to examine the potential of quantum kernels to extrapolate. 

Figure 7 compares the extrapolation accuracy of quantum models with both entangled and untangled kernels and the classical model with the RBF kernel. 
The results shown in Figure 7 are obtained with models trained by random samples of {\it ab initio} potential energy points from the energy interval below the energy threshold indicated on the horizontal axis. 
The RMSEs shown are calculated for the entire energy range of the PES extending to 21,000 cm$^{-1}$. Figure 7 illustrates two important results. First, including the entanglement between qubits into the quantum circuit enhances the extrapolation accuracy to a great extent. Second, models 
with entangled kernels appear to outperform models with the RBF kernels for low thresholds of the training data range, corresponding to a larger extrapolation interval. 

To illustrate the comparison between the model predictions and the original {\it ab initio} energies, we show in Figure 8 the results of several models corresponding to different energy ranges of the training samples (shown by the shaded intervals). All models illustrated in Figure 8 are trained by 1500 {\it ab initio} points. 
The lines represent the GP model predictions and the symbols --   the potential energy points sampled as functions of the separation between H$_2^+$ and OH fragments.
As in Figure 4, at each value of $R$, we locate the energy point in the original set of {\it ab initio} points by varying the angles and/or the interatomic distances within the fragments. This energy point is then compared with the GP predictions.
The functional form of PES at high energies is qualitatively different from that at low energies. 
Figure 8 shows that optimized quantum kernels can produce GP models that generalize predictions to different function distributions.

\section{Conclusion}

We have demonstrated that quantum circuits of gate-based quantum computers can be used to build kernels for regression models of global PES for polyatomic molecules. 
Such kernels can be obtained by measuring the individual qubit states.
We have shown that such kernels can be constructed with a fixed quantum circuit ansatz, previously used for classification problems, 
provided the quantum gate parameters are optimized to maximize $\log [{\cal L} + 1]$, where ${\cal L}$ is marginal likelihood. 
This yields Gaussian process models of PES with quantum kernels. 
While the standard procedure for training Gaussian process models is to maximize $\log {\cal L}$, our results illustrate that $\log {\cal L}$ is very sensitive to variation of the circuit parameters, making the optimization challenging. However, we have shown that maximization of $\log [{\cal L} + 1]$ can be performed with Bayesian optimization, 
yielding stable results that correspond to accurate regression models with quantum kernels.

We have compared the accuracy of Gaussian process models of PES with quantum kernels based on entangled qubits, quantum kernels with unentangled kernels and classical Gaussian process models with RBF kernels. 
In all cases considered, the accuracy of quantum models including two-qubit rotation gates is comparable with the accuracy of classical models with RBF kernels.  
The quantum models with entangled kernels outperform the classical models with optimized RBF kernels for the class of problems aiming to construct the 6D PES at high energies based on 1500  {\it ab initio} points at low energies.
At the same time, the accuracy of all quantum models drops significantly, when the entangling two-qubit gates are omitted from the quantum circuits. This illustrates the critical role of qubit entanglement in the quantum kernel computation algorithm. 

Our work demonstrates that quantum kernels obtained with a small number of qubits and quantum gates can be used for accurate regression models. 
This is important because finite fidelity of current NISQ devices is a major obstacle to increasing the size of quantum circuits.  The quantum circuit used in the present work can be readily implemented on the current IBM quantum computer. 
Moreover, we have built quantum kernels for Gaussian process models, 
which themselves could be used as surrogate models underlying Bayesian optimization.  Thus, our work complements Ref. \cite{qgp} to pave the way for the development of the quantum analogue of Bayesian optimization. 
If quantum kernels prove to offer better inference for supervised learning tasks with a small number of training points than classical kernels, Bayesian optimization with quantum GPs may offer a useful application of quantum computing to optimization of functions that are exceedingly expensive to evaluate. 

\section*{acknowledgment}
This work was supported by NSERC of Canada



\clearpage
\newpage

\begin{figure}[http]
\centering
\includegraphics[scale=0.30]{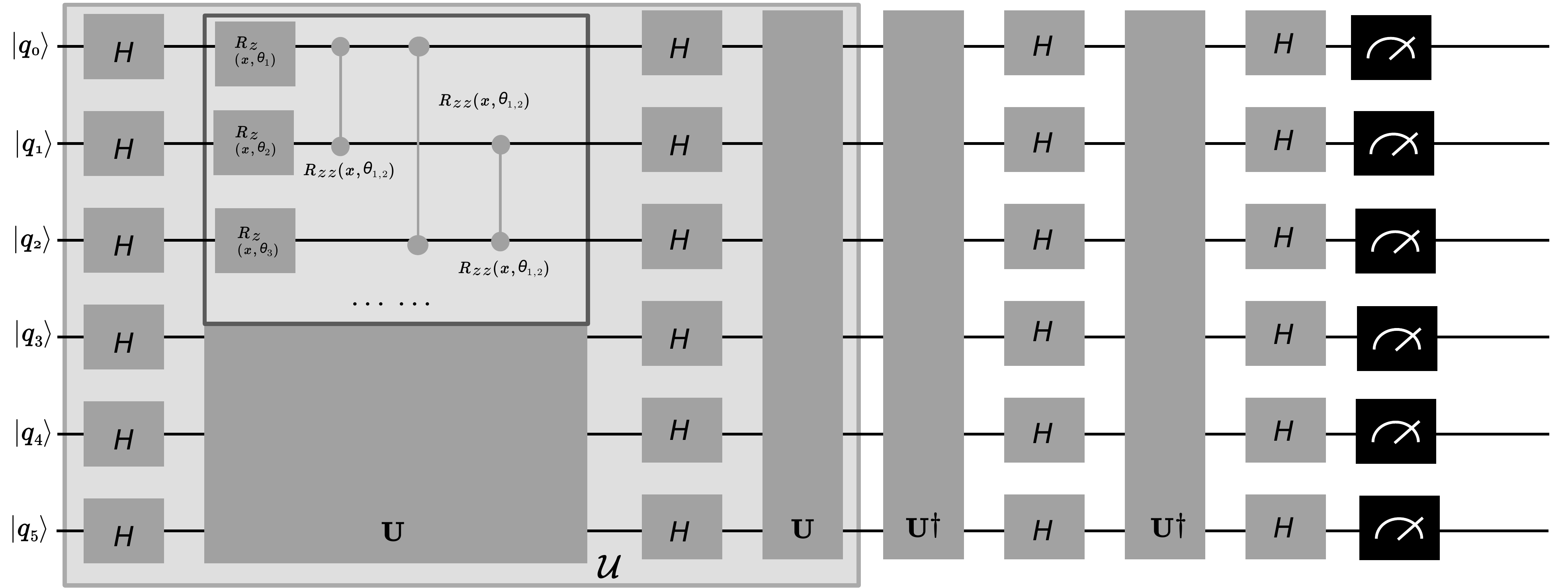} 
\caption{Quantum circuit used in the present work to build quantum kernels of Gaussian process models. 
The sequence of gates in $\bf U$ is determined by Eq. (\ref{circuit}). $H$ denotes Hadamard gates and $R_Z$ -- single qubit rotation gates. See text for more details. 
}
\label{fig:circuits}
\end{figure}

\begin{figure}[http]
\centering
\includegraphics[scale=0.5]{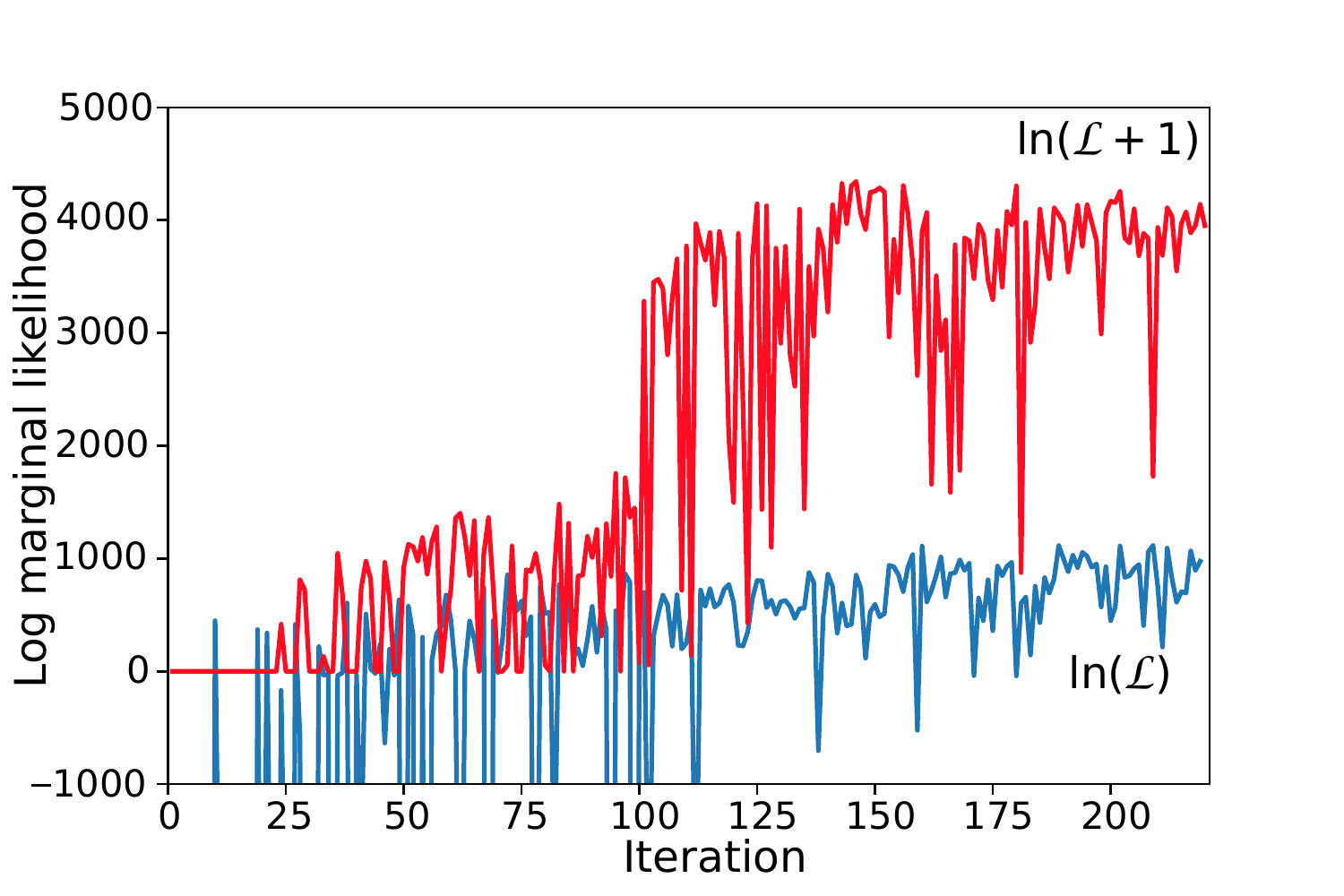} 
\includegraphics[scale=0.5]{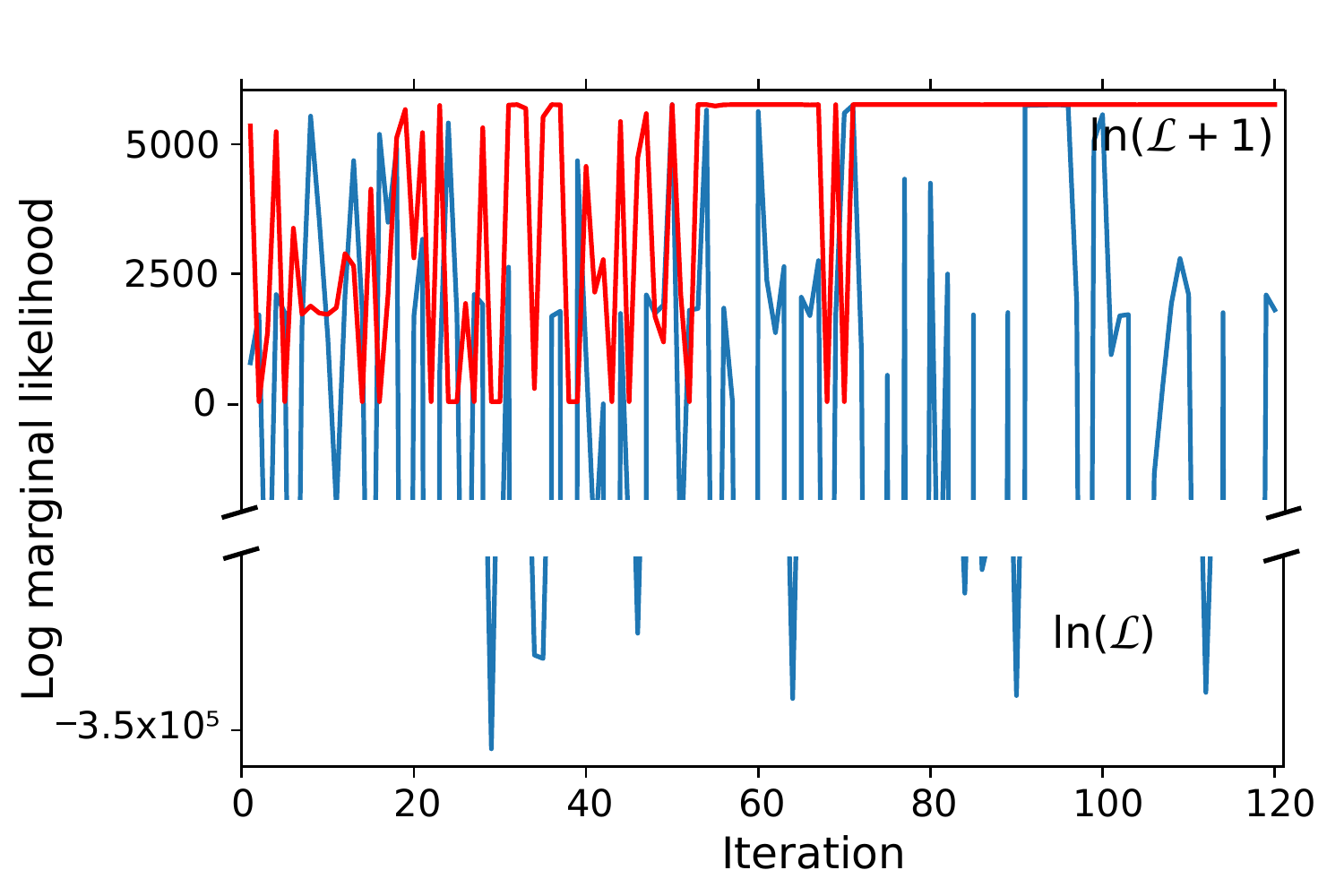}
\caption{LML of quantum GP models with unentangled kernels (left panel) and entangled kernels (right panel) as functions of the number of BO iterations. 
 Upper curves (red): LML obtained by maximization of ${\cal O}$ as defined by Eq. (\ref{o-function}).
Lower curves (blue): LML obtained by maximization of $\log {\cal L}$.
All GPs are trained by the same set of $n=1000$ energy points randomly selected from the entire energy range $[0,21000] \, \mathrm{cm}^{-1}$.}
\label{fig:ln(L+1)}
\end{figure}

\begin{figure}[http]
\centering
\includegraphics[scale=0.5]{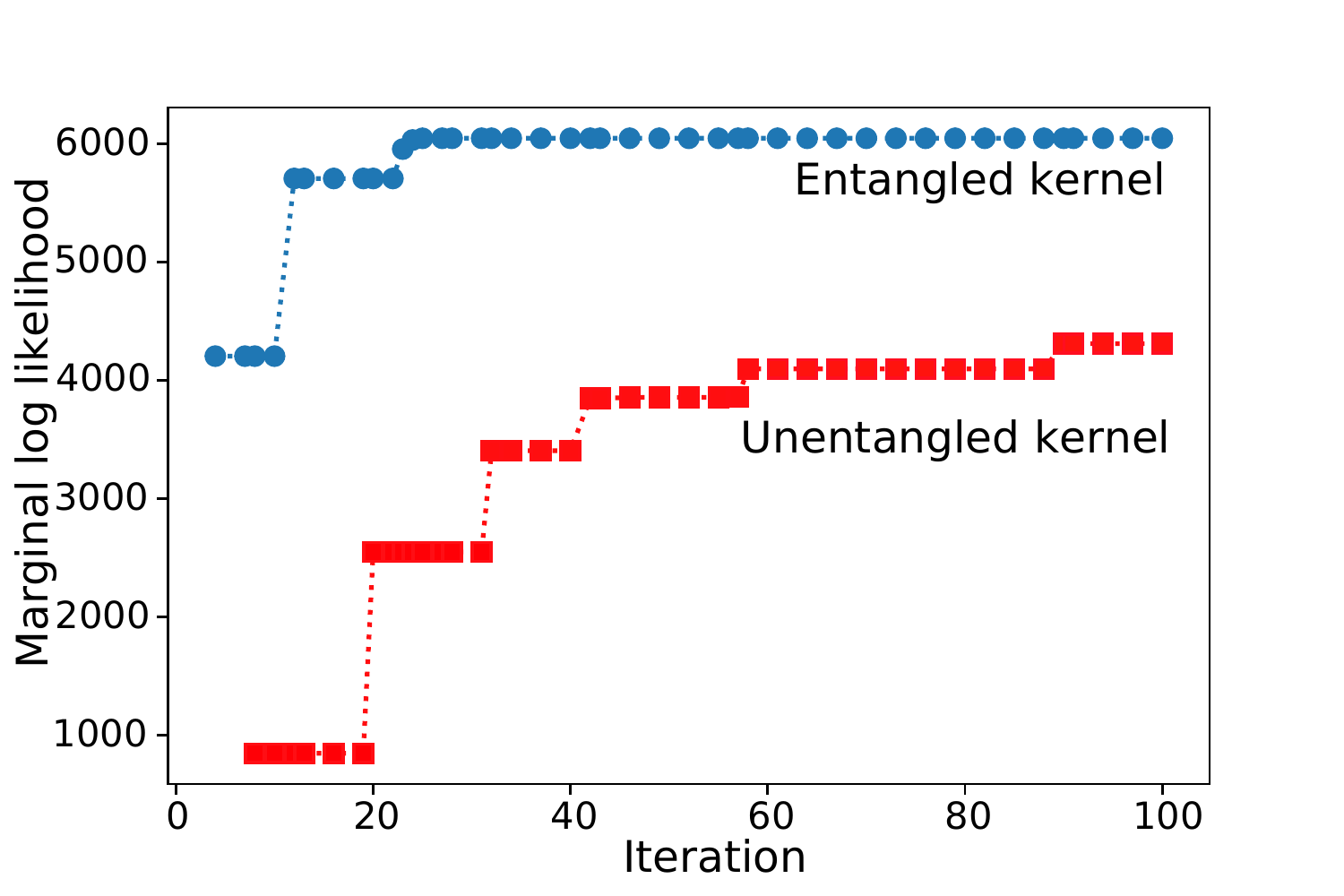} 
\includegraphics[scale=0.5]{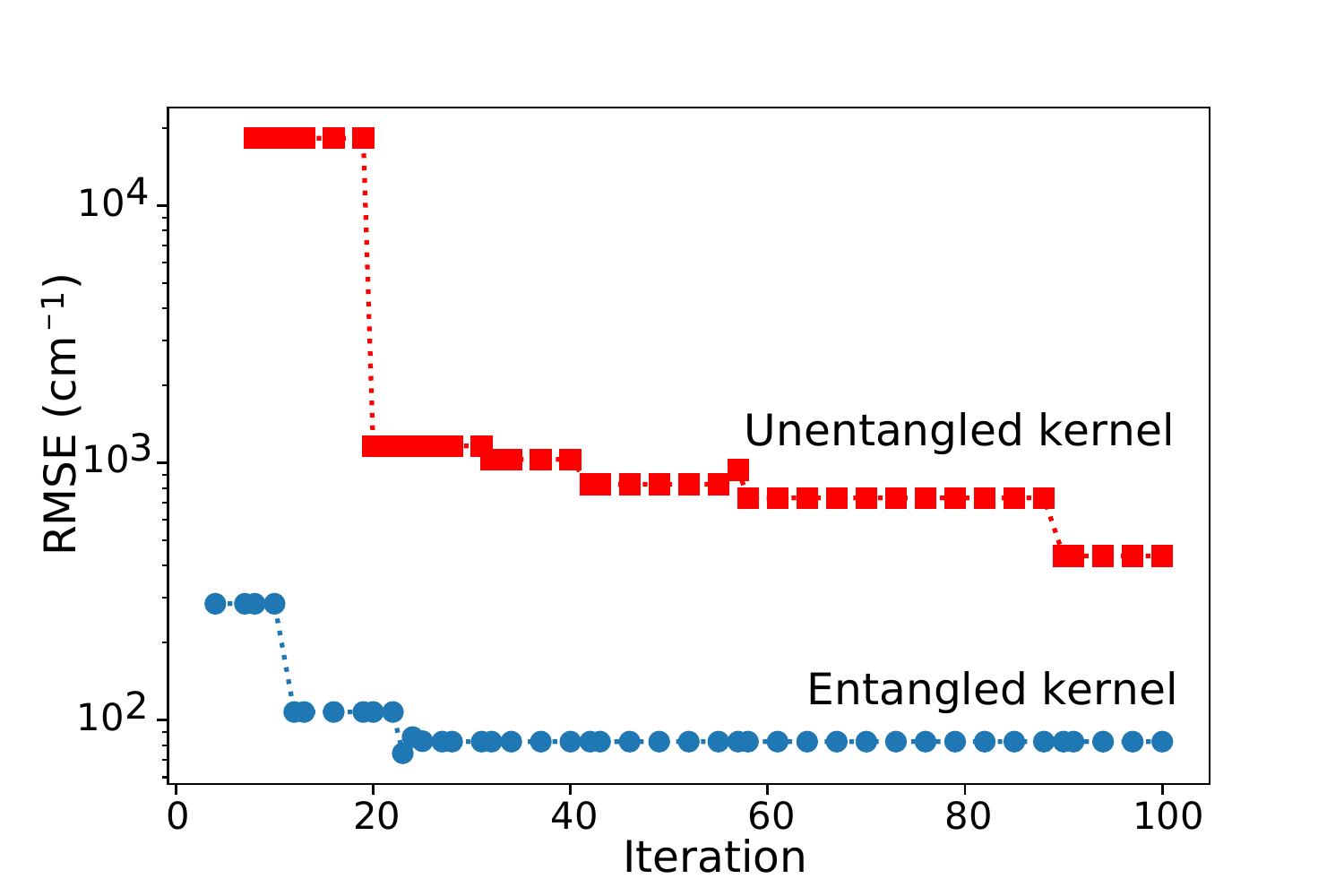}
\caption{Left panel: Maximum value of LML (left panel) and RMSE (right panel) of quantum GP models with entangled and unentangled kernels as functions of the number of BO iterations.
All GPs are trained by the same set of $n=1000$ energy points randomly selected from the entire energy range $[0,21000] \, \mathrm{cm}^{-1}$.
}
\label{fig:BO_entangle}
\end{figure}

\begin{figure}[http]
\centering
\includegraphics[scale=0.5]{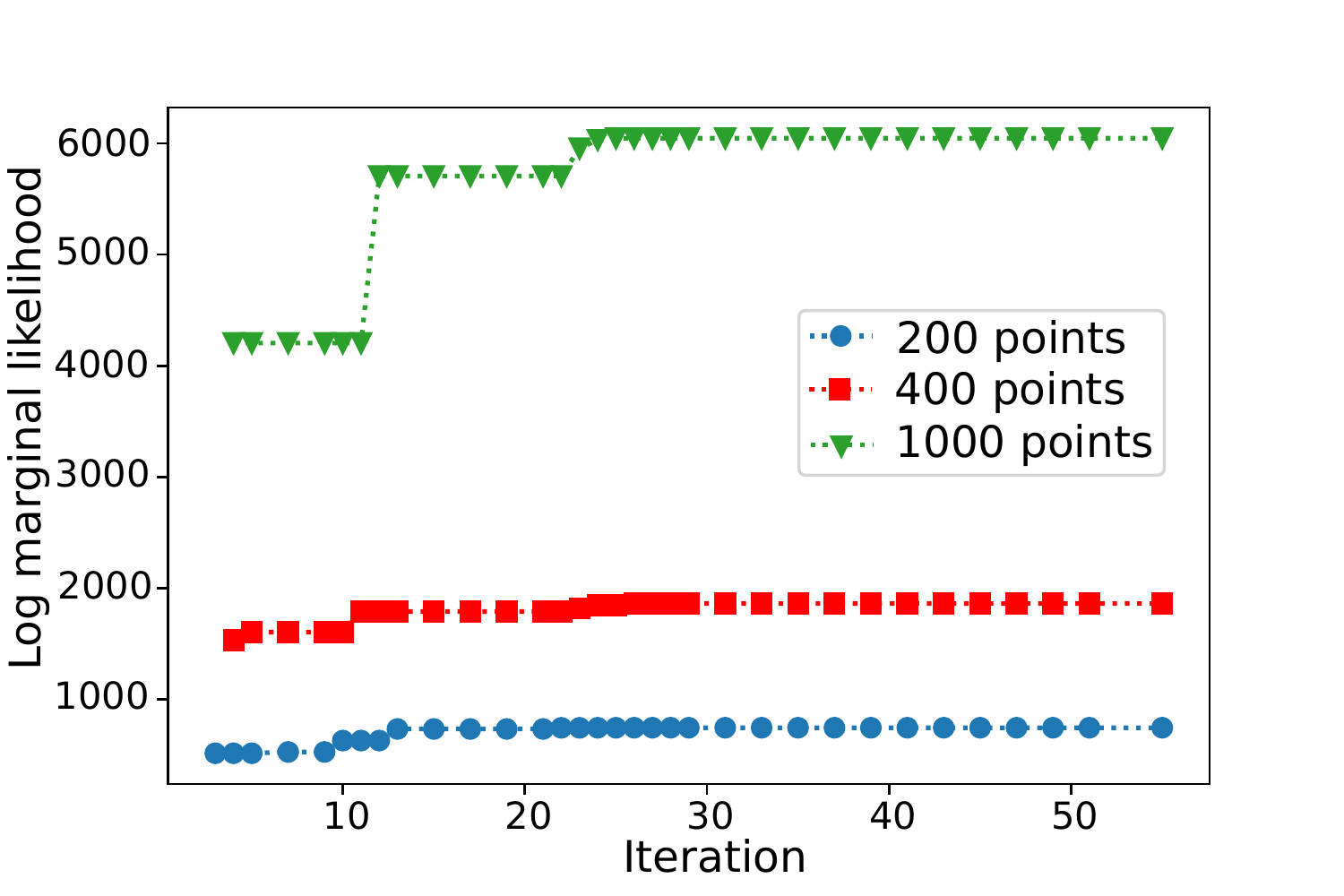} 
\includegraphics[scale=0.5]{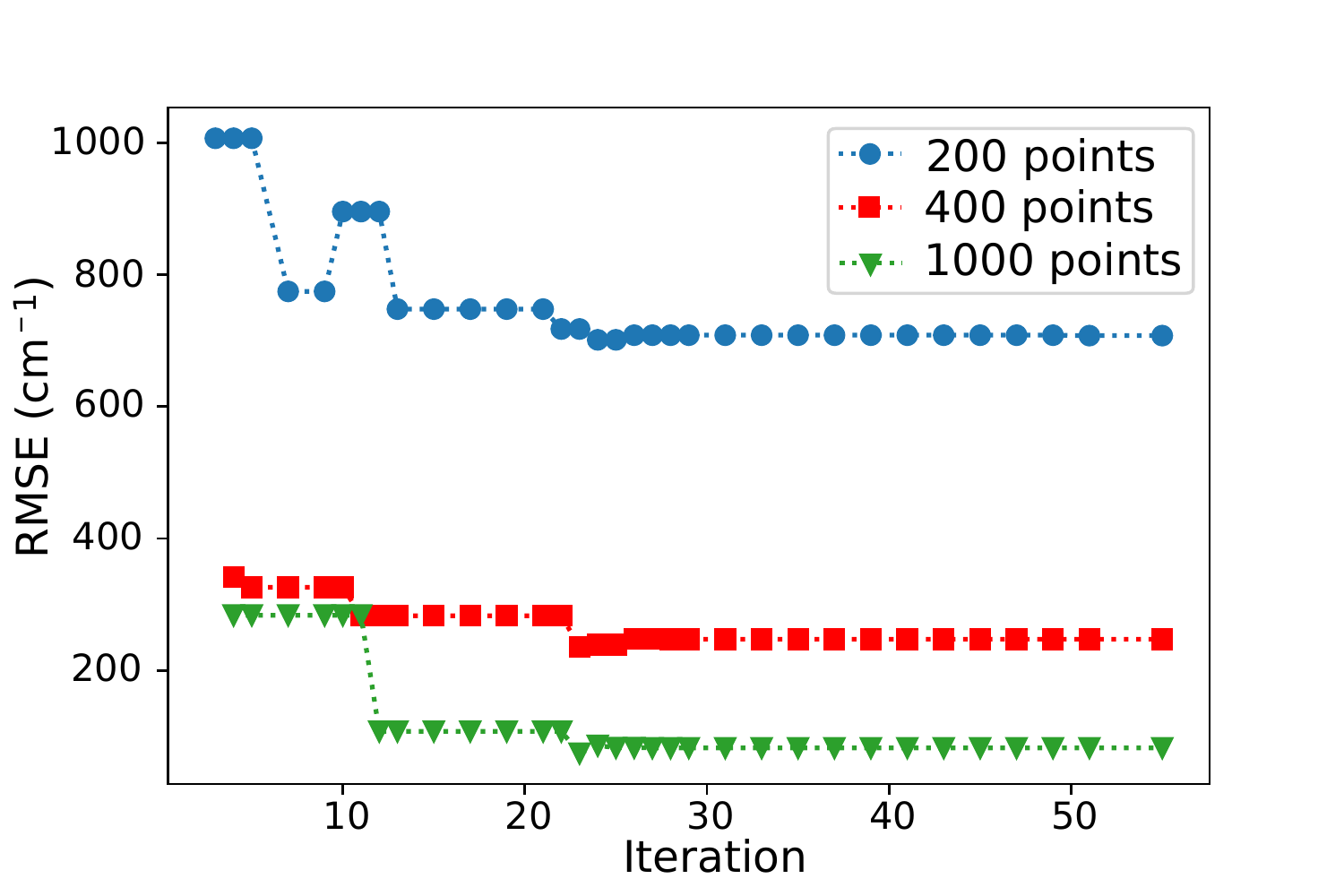}
\caption{
Maximum value of LML (left panel) and RMSE (right panel) of quantum GP models with entangled kernels as functions of the number of BO iterations for different numbers of training points:
circles -- $n = 200$; squares -- $n = 400$; triangles -- $n = 1000$. The models are trained by {\it ab initio} points randomly sampled from  
the energy interval $[0,21000] \, \mathrm{cm}^{-1}$.}
\label{fig:BO}
\end{figure}

\begin{figure}[http]
\centering
\includegraphics[scale=0.8]{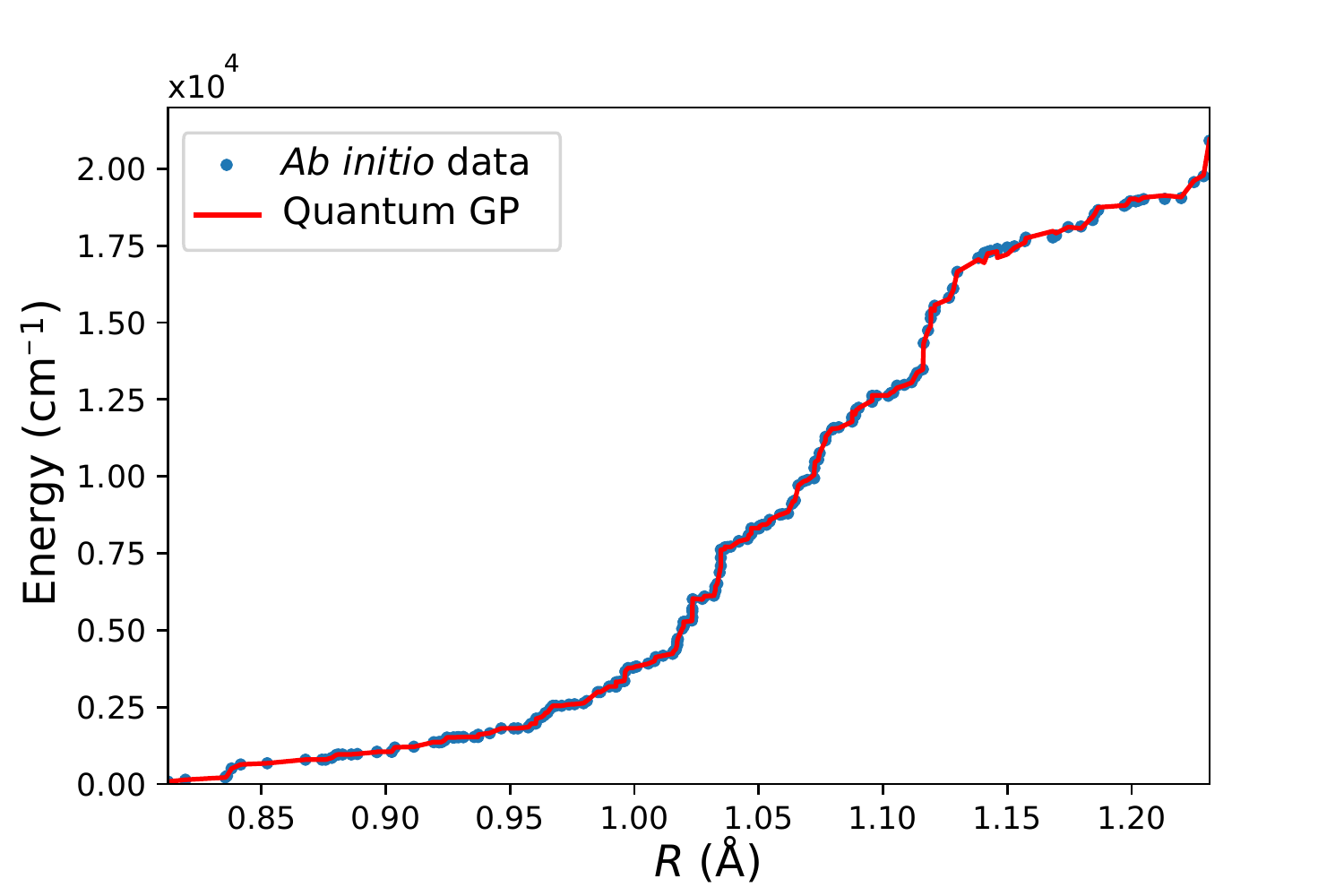} 
\caption{Comparison of quantum GP model predictions (solid curve) with the original potential energy points (symbols) for H$_3$O$^+$ as functions of the separation between  the H$_2^+$ and OH fragments. 
The variable R specifies the distance between the O atom and one of the H atoms in the H$_2^+$ fragment. At each value of $R$, we locate the energy point in the original set of {\it ab initio} points by varying the angles and/or the interatomic distances within the fragments. This energy point is then compared with the GP predictions. 
The 6D GP model is trained by 1000 {\it ab initio} points randomly selected from the entire energy range and uses the entangled kernel.}
\label{fig:RMSE}
\end{figure}

\begin{figure}[http]
\centering
\includegraphics[scale=0.8]{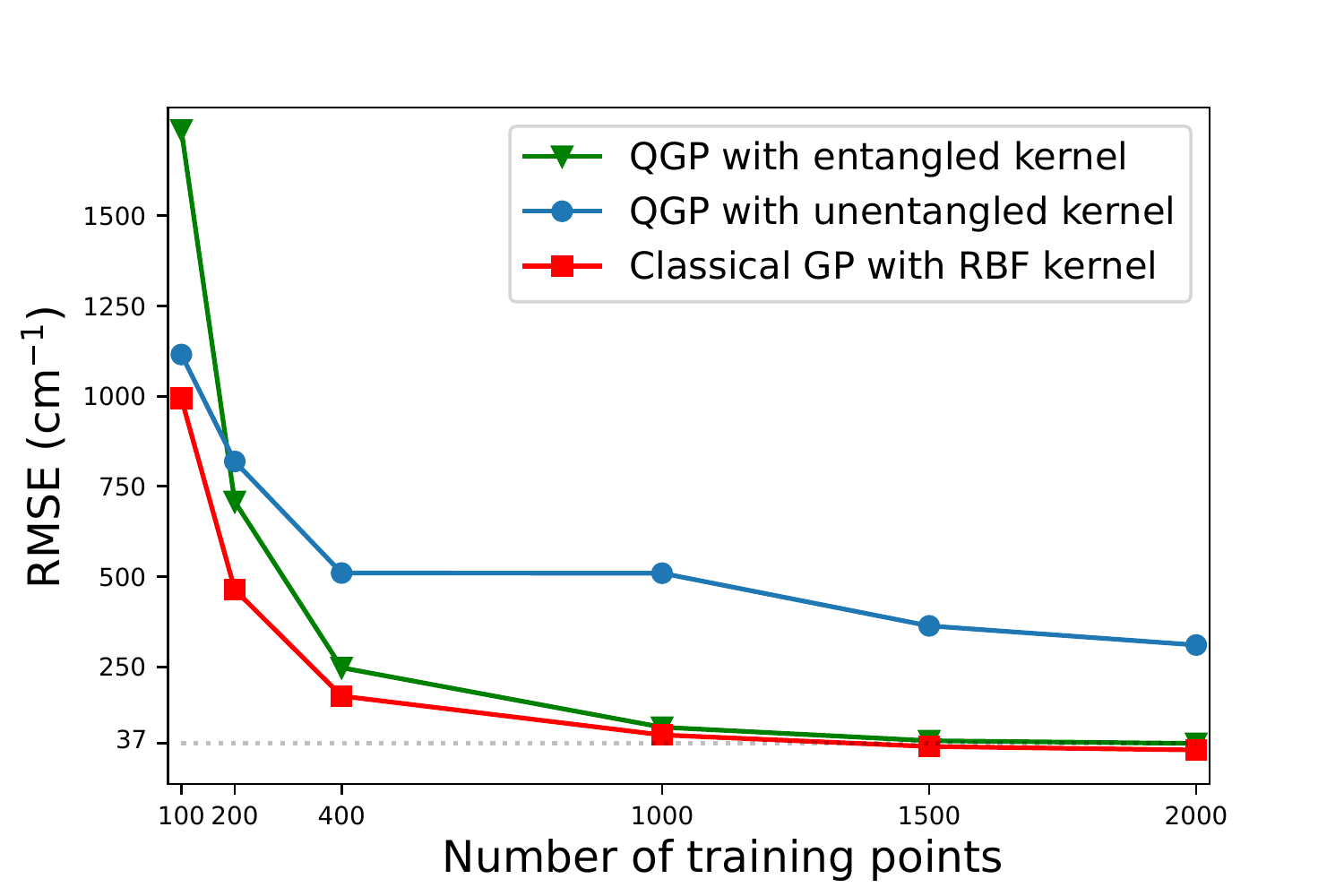} 
\caption{Dependence of the RMSE for GP models with quantum kernels based on quantum circuits with unentangled qubits (triangles),  entangled qubits (circles) and classical RBF kernel (stars) on the number of training energy points.
The models are trained by {\it ab initio} points randomly sampled from  
the energy interval $[0,21000] \, \mathrm{cm}^{-1}$.
The RMSEs are calculated using all remaining energy points in the same energy interval that are not used for training. }
\label{fig:RMSE_NUM}
\end{figure}

\begin{figure}[http]
\centering
\includegraphics[scale=0.8]{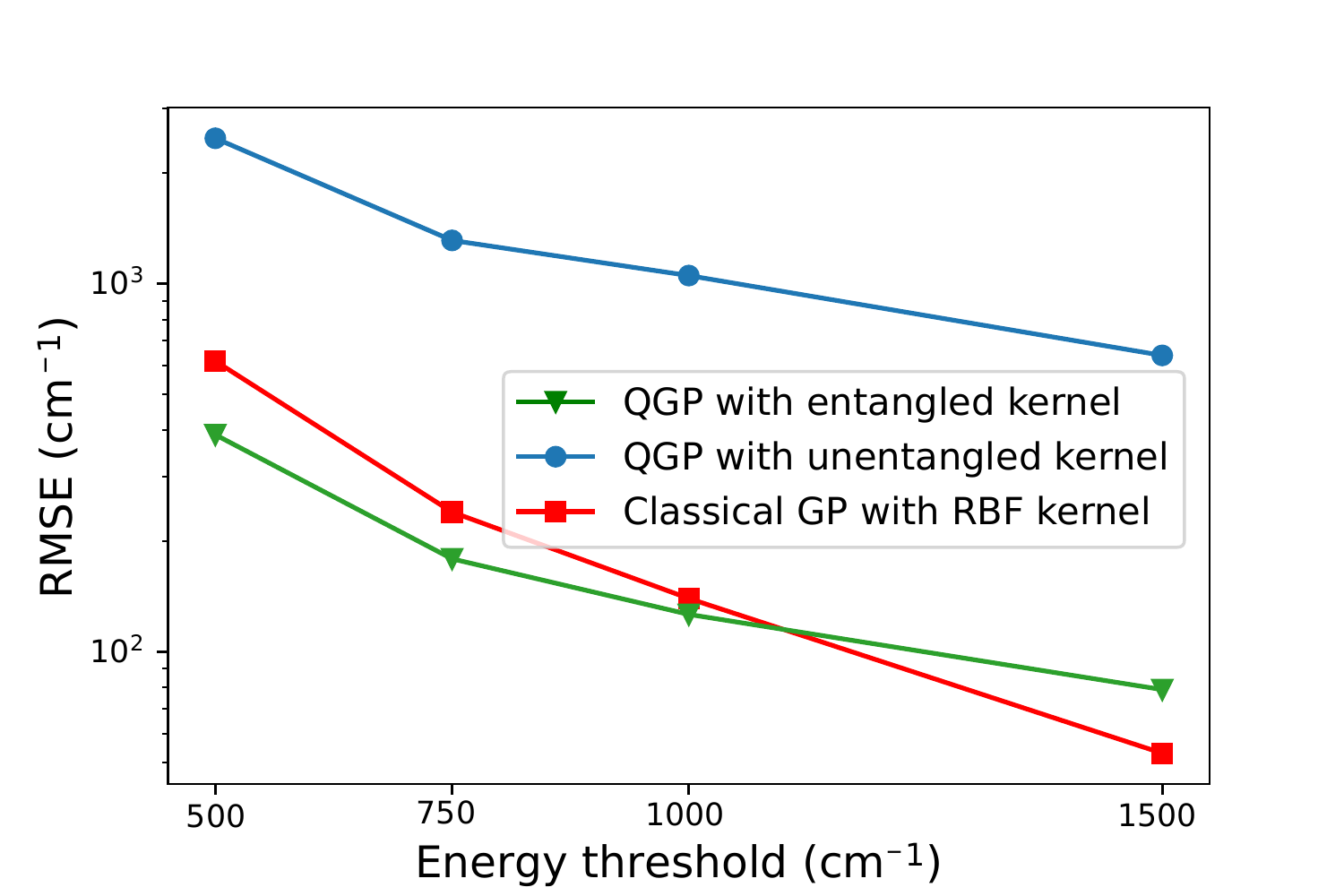} 
\caption{Extrapolation in the energy domain: RMSE for GP models with quantum kernels based on quantum circuits with unentangled qubits (triangles),  entangled qubits (circles) and classical RBF kernel (stars) as functions of the training energy threshold. 
All models are trained by 1500 randomly selected {\it ab initio} points from the energy interval below the indicated energy threshold. The RMSEs are calculated using all remaining energy points that are not used for training and that cover the energy interval $[0,21000] \, \mathrm{cm}^{-1}$.
}
\label{fig:RMSE_extrapolation}
\end{figure}

\newpage

\begin{figure}[http]
\centering
\includegraphics[scale=0.5]{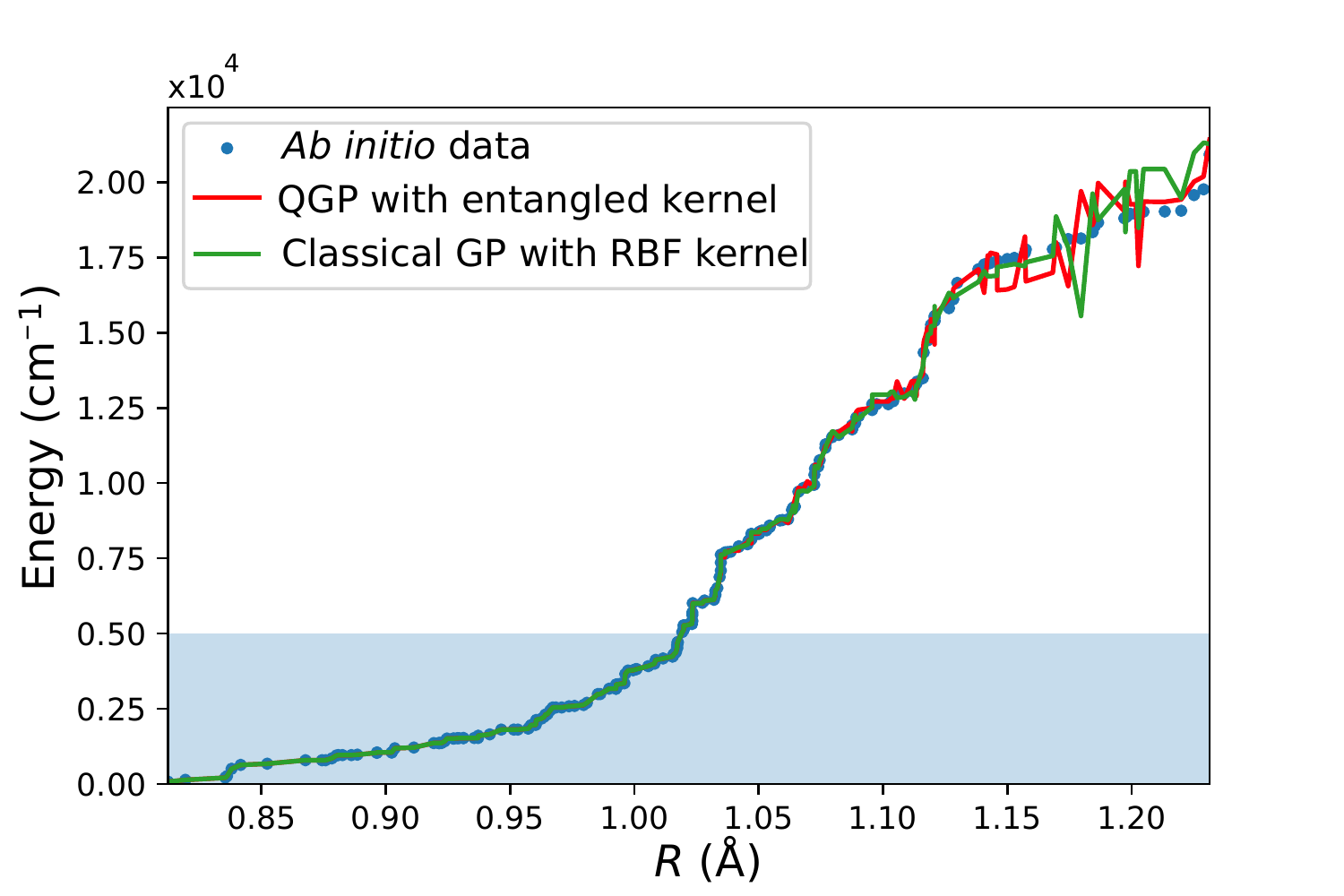} 
\includegraphics[scale=0.5]{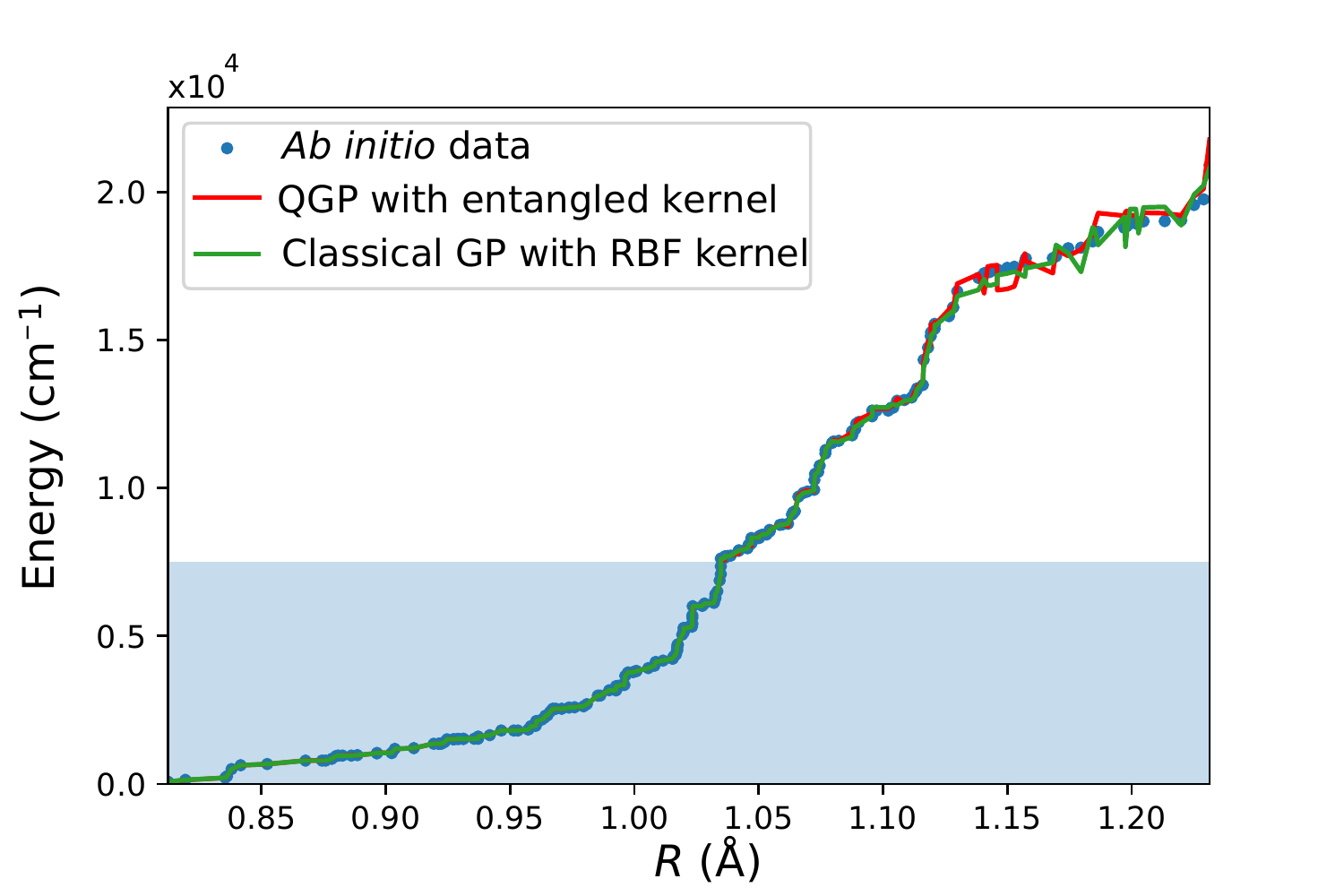}
\includegraphics[scale=0.5]{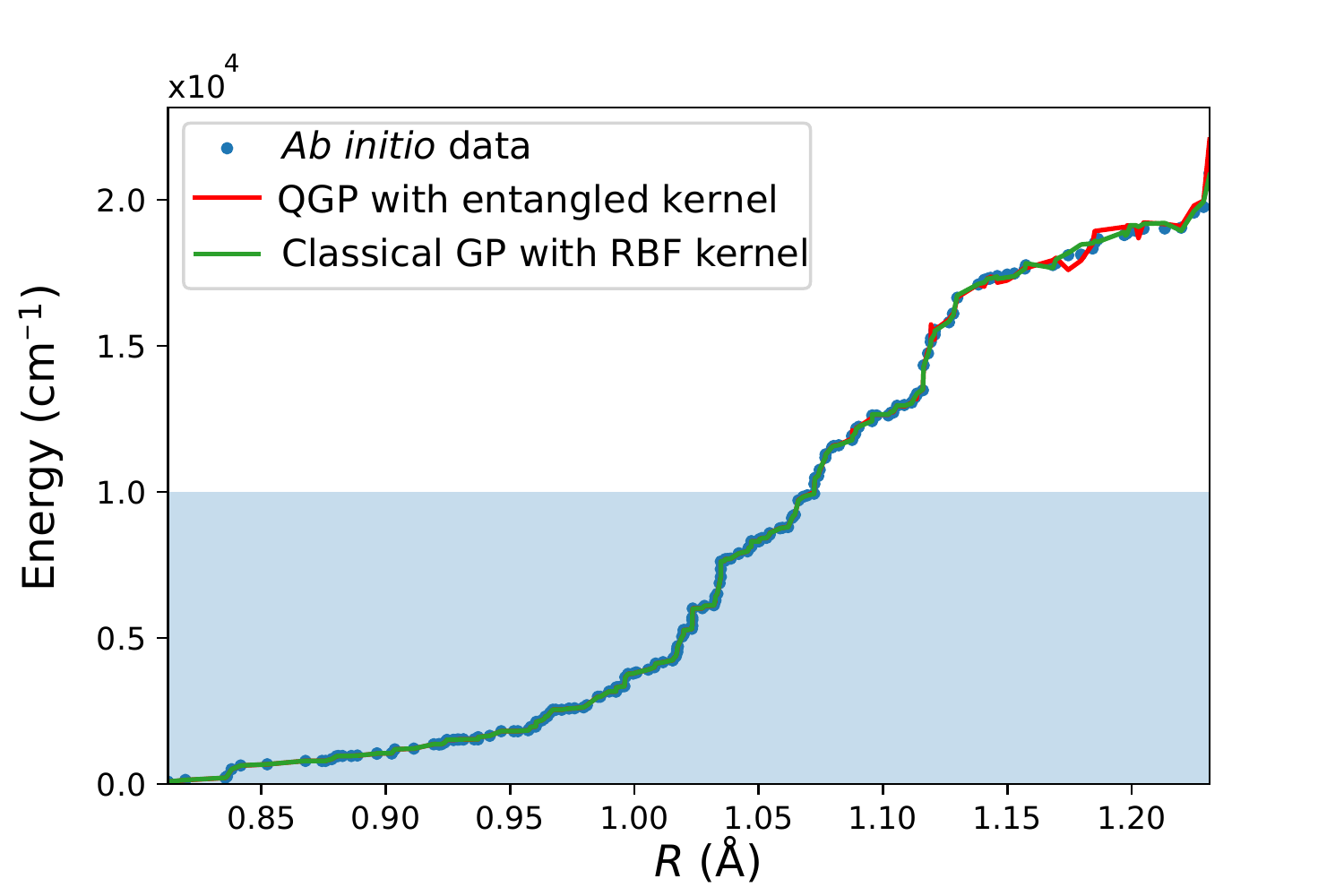}
\includegraphics[scale=0.5]{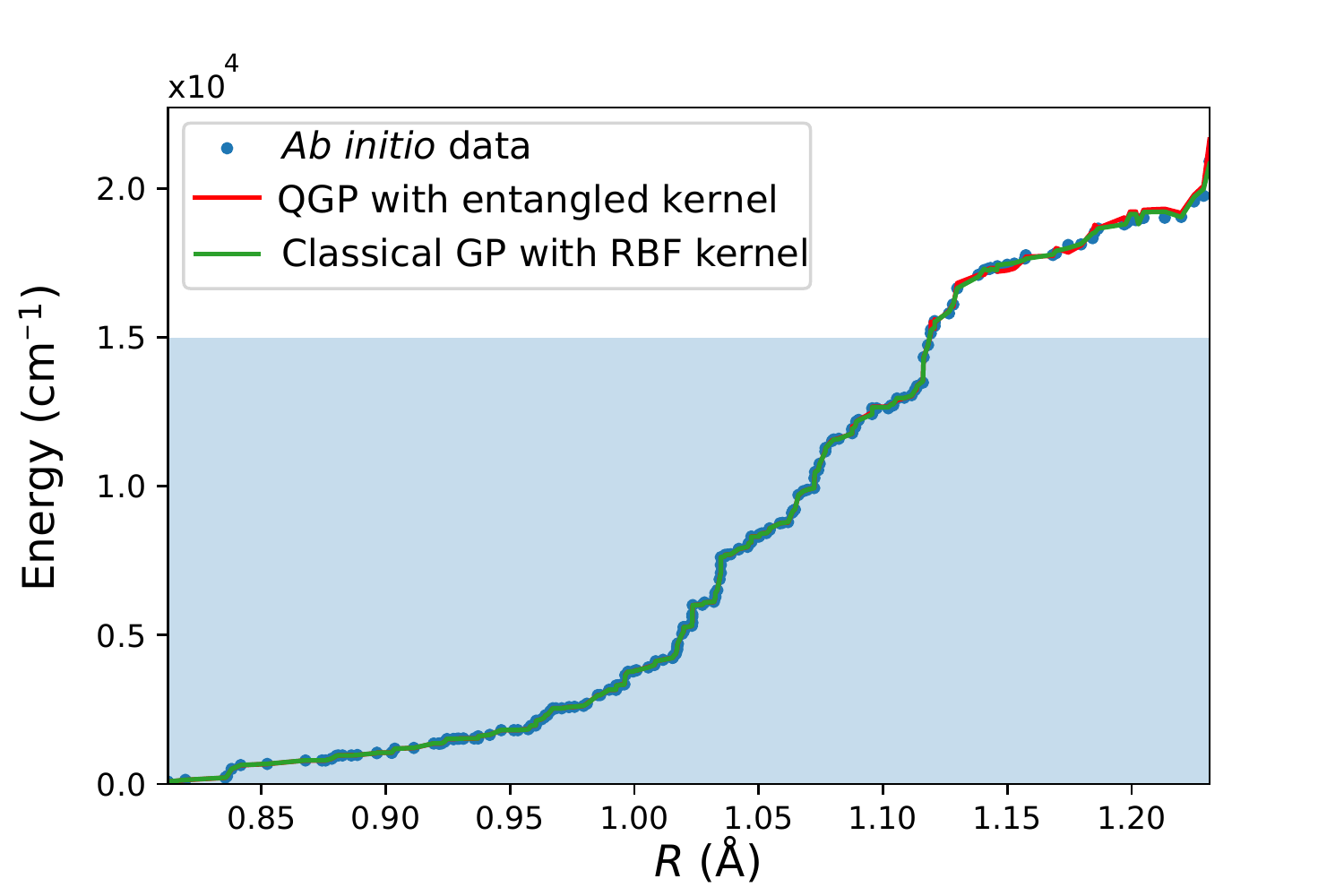}
\caption{Comparison of GP model predictions (solid curve) with the original potential energy points (symbols) for H$_3$O$^+$ as functions of the separation between  the H$_2^+$ and OH fragments.
The GP models are trained by 1500  {\it ab initio points} randomly selected from the energy interval shown by the blue shaded region. 
The variable $R$ specifies the distance between the O atom and one of the H atoms in the H$_2^+$ fragment. 
At each value of $R$, we locate the energy point in the original set of {\it ab initio} points by varying the angles and/or the interatomic distances within the fragments. This energy point is then compared with the GP predictions.}
\label{fig:Extrapolation}
\end{figure}


\newpage


\end{document}